\documentclass[aps,prb,twocolumn,showpacs,superscriptaddress]{revtex4-1}

\usepackage{graphicx} 

\begin{document}

\title
{Armchair nanoribbons of silicon and germanium honeycomb structures}

\author{S. Cahangirov}
\affiliation{UNAM-Institute of Materials Science and
Nanotechnology, Bilkent University, Ankara 06800, Turkey}
\author{M. Topsakal}
\affiliation{UNAM-Institute of Materials Science and
Nanotechnology, Bilkent University, Ankara 06800, Turkey}
\author{S. Ciraci} \email{ciraci@fen.bilkent.edu.tr}
\affiliation{UNAM-Institute of Materials Science
and Nanotechnology, Bilkent University, Ankara 06800, Turkey}
\affiliation{Department of Physics, Bilkent University, Ankara
06800, Turkey}

\date{\today}

\begin{abstract}
We present a first-principles study of bare and hydrogen
passivated armchair nanoribbons of the puckered single layer
honeycomb structures of silicon and germanium. Our study includes
optimization of atomic structure, stability analysis based on the
calculation of phonon dispersions, electronic structure and the
variation of band gap with the width of the ribbon. The band gaps
of silicon and germanium nanoribbons exhibit family behavior
similar to those of graphene nanoribbons. The edges of bare
nanoribbons are sharply reconstructed, which can be eliminated by
the hydrogen termination of dangling bonds at the edges. Periodic
modulation of the nanoribbon width results in a superlattice structure which
can act as a multiple quantum wells. Specific electronic states
are confined in these wells. Confinement trends are qualitatively
explained by including the effects of the interface. In order to
investigate wide and long superlattice structures we also
performed empirical tight binding calculations with parameters
determined from \textit{ab initio} calculations.
\end{abstract}

\pacs{73.22.-f, 63.22.-m, 68.65.-k} \maketitle

\section{Introduction}
Graphene, a two-dimensional (2D) honeycomb structure of
single layer of carbon atoms, have been synthesized and
established as a material with wide range 
of unusual properties.\cite{novo,zhang,berger}
It is a semimetal with a band profile having linear dispersion
near the Fermi level,\cite{dirac} which attributes to its
electrons a massless Dirac Fermion behavior. Quasi one-dimensional
(1D) derivatives of graphene, called graphene nanoribbons, were also
produced recently.\cite{barbaros,dai} Graphene nanoribbons are
semiconductors having interesting electronic properties depending
on their geometry. These properties can be used to fabricate
nanodevices as field effect transistors, spin valves, multiple
quantum wells and
etc.\cite{novo_nmats,dai_transistor,cohalf,metosa,hal}

2D honeycomb structures and 1D nanotubes of Si and Ge
were studied earlier.\cite{takeda,durgun} Stringent stability
tests have recently shown that, 2D honeycomb structures of Si and
Ge can be found stable in a slightly buckled geometry.\cite{prl}
These structures have similar properties as graphene and thus
carry the potential of being used in the similar
applications.\cite{nature} Compared to graphene, the interatomic
distance is larger in Si and Ge, so the diminished $\pi-\pi$
overlaps cannot maintain the planar stability anymore. Eventually,
the $sp^2$ hybrid orbitals are slightly dehybridized to form
$sp^3$-like orbitals, which in turn results in a puckered
structure.\cite{ency} Freestanding graphene sheets and nanoribbons
can be produced spontaneously, but it is not the case for 2D
honeycomb structures of Si (silicene) and Ge. However, there are
plenty experimental work on growth of Si nanoribbons especially on
Ag surface.\cite{nakamura,kara} These highly metallic nanoribbons
are formed by self-organization and have straight, atomically
perfect and massively parallel structures.\cite{kara} The
electronic structure of Si nanoribbons on Ag surface was also
investigated theoretically.\cite{karacomp}

This paper reveals the atomic and electronic structure of armchair
Si and Ge nanoribbons. Stability analysis, based on the
calculation of the phonon dispersions via the force constant
method, was performed. Energy band structure calculated by
first-principles density functional theory (DFT) were used to
generate the parameters of the tight-binding model. It is
found that armchair nanoribbons of Si and Ge are stable and their
band gap vary with their width displaying a family
behavior.\cite{gaps,gapsgw} Formation of multiple quantum well
structure in superlattices consisting of periodically repeated
junctions of nanoribbons having different widths was also
investigated in detail. A 1D model was proposed to
understand the effect of the interface in superlattice structures. 
It is found that specific electronic states are confined in these 
superlattice structures. The interface effects, which can explain 
unexpected confinements, are revealed.

\section{Methods}
We have performed first-principles plane wave calculations within 
Local Density Approximation (LDA) \cite{ceperley} using projector
augmented wave (PAW) potentials \cite{blochl}. All structures are
treated within supercell geometry using the periodic boundary
conditions. A plane-wave basis set with kinetic energy cutoff of
300~eV is used. In the self-consistent potential and total energy
calculations, the Brillouin zone (BZ) is sampled by ($15 \times 1
\times 1$) special \textbf{k}-points. All atomic positions and
lattice constants are optimized by minimization of the total
energy and atomic forces. The vacuum separation between the
nanoribbons in the adjacent unit cells is taken to be at least
10~\AA. The convergence for energy is chosen as 10$^{-5}$~eV
between two steps, and the maximum Hellmann-Feynman forces acting
on each atom is less than 0.02 eV/\AA{} upon ionic relaxation.
Numerical plane wave calculations have been performed by using
VASP package.\cite{vasp1,vasp2} Phonon dispersions were obtained
using the force constant method with forces calculated in a ($5
\times 1 \times 1$) supercell.\cite{kfh,alfe} Coupling parameters
of empirical tight binding calculations are determined from the
\textit{ab initio} results and are used to treat wide and long
superlattice structures, comprising as many as 3600 Si atoms.

\begin{figure}
\includegraphics[width=8.4cm]{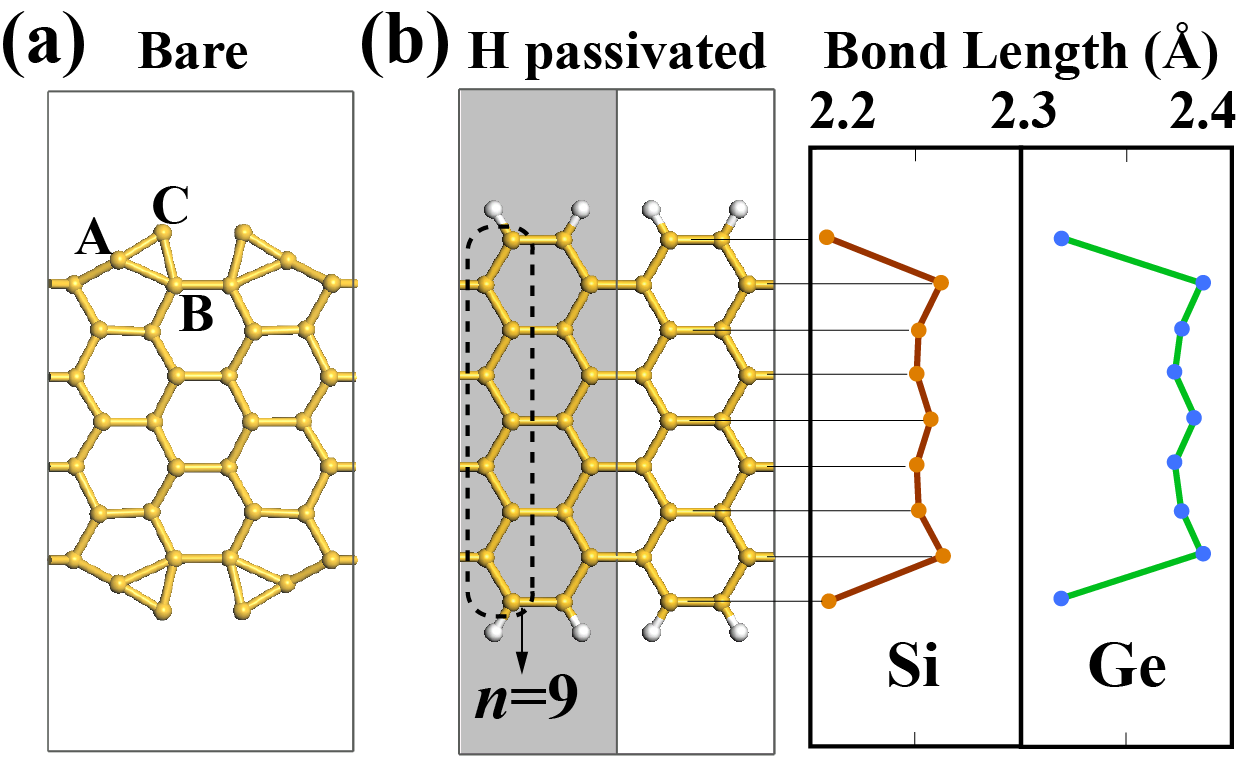}
\caption{(a) Atomic structure of fully relaxed bare armchair Si
nanoribbon of width $n=9$ (ASiNR-9). (b) Atomic structure and bond
length distribution of hydrogen saturated  ASiNR-9. Similar
pattern is observed in hydrogen saturated AGeNR-9. The primitive 
unit cell of ASiNR-9 is shaded. The zigzag chain of Si atoms 
perpendicular to the nanoribbon axis is delineated by the dashed lines.}
\label{fig:figa}
\end{figure}

\section{Atomic Structure and Stability}
We first investigate the atomic structure of bare and hydrogen
saturated armchair nanoribbons of Si and Ge. The ideal honeycomb
structure is cut parallel to the nearest neighbor bonds to form an
ideal bare nanoribbon with a certain width. Armchair nanoribbons are
classified by counting the number, $n$, of Si (or Ge) atoms forming 
a zigzag chain perpendicular to the cut direction. Accordingly, there 
are $2n$ Si (or Ge) atoms in the primitive unit cell of an ideal bare 
armchair nanoribbon. This structure is treated by a supercell 
having periodic boundary condition in cut
direction and a vacuum spacing in other directions. To lift the
constraints imposed by (1$\times$1) unit cell, we have used
(2$\times$1) supercell. Figure \ref{fig:figa}(a) presents the
atomic structure of a sample bare armchair Si nanoribbon after
structural relaxation. Here one can see a (2$\times$1)
reconstruction at the edges, which would be missed if (1$\times$1)
unit cell was used in the calculation. The edge reconstruction
is reminiscent of the reconstruction of Si(100)-(2$\times$1) surface. In 
the latter case, two adjacent surface Si atoms each having two $sp^3$-dangling bonds
come closer and form a new dimer bond using one $sp^3$-dangling
bond from each atom. At the end, the number of $sp^3$ dangling
bonds is halved and hence the energy is lowered through reconstruction. Similarly, in
Fig.\ref{fig:figa}(a) A and B atoms come closer to form a bond.
Since the nature of bonding is modified around B atom, the ABC
triangle is bowed. At the end, the number of the $sp^2$ dangling bonds
is halved. Note that, this kind of reconstruction is not seen in
bare armchair graphene nanoribbons. Moreover, in graphene and its
nanoribbons all atoms lie in the same plane, while structures
considered here are slightly buckled. The separation between
adjacent atoms in the perpendicular direction to the plane is
around 0.4~\AA~and 0.6~\AA~for Si and Ge nanoribbons,
respectively.

Figure \ref{fig:figa}(b) presents the atomic structure and bond
length distribution of a sample hydrogen saturated armchair
silicon nanoribbons. In contrast to bare nanoribbons, saturation
by hydrogen lifts the (2$\times$1) reconstruction at the edges.
In Fig.~\ref{fig:figa}(b) there are $n=9$ Si atoms forming zigzag 
chain perpendicular to the nanoribbon axis and hence this 
armchair nanoribbon is classified as ASiNR-9. Accordingly 
the number of Si (or Ge) atoms in the primitive unit cell is $2n$. Note
that the bond length distribution is nearly uniform except a
sudden decrease at the edges. This pattern was also observed in
armchair graphene nanoribbons.\cite{gaps}

\begin{figure}
\includegraphics[width=8.4cm]{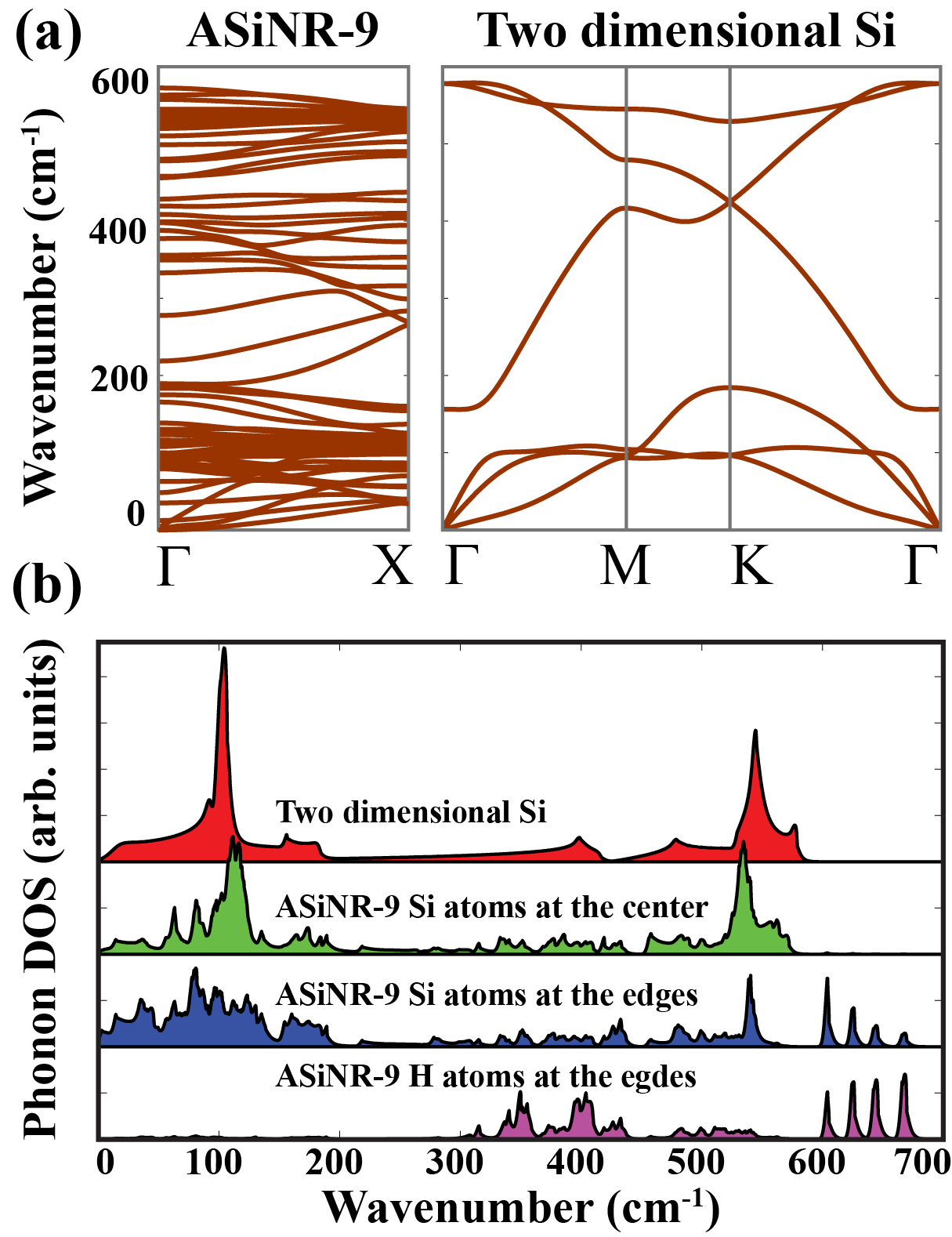}
\caption{(a) Phonon dispersions calculated for hydrogen saturated ASiNR-9 and
for 2D honeycomb structure of Si. States that appear above 
600~cm$^{-1}$ are related to Si-H bonds of hydrogen saturated ASiNR-9
and were not shown. (b) Phonon densities of states (DOS) of the
hydrogen saturated ASiNR-9 projected to Si atoms at the center and
at the edges and also to H atoms at the edges. DOS of 2D
honeycomb structure of Si is also presented for comparison.}
\label{fig:figb}
\end{figure}

Figure \ref{fig:figb}(a) presents the phonon dispersion profile
for hydrogen saturated ASiNR-9. Also phonon dispersion profile of
2D silicon, reproduced from Ref.~[\onlinecite{prl}], was shown for
comparison. It is not possible to generate the dispersion profile
of ASiNR by folding that of 2D Si, but the pattern is similar to
that found in nanotube dispersion profiles obtained by
zonefolding.\cite{dubay} In the phonon dispersion profile of
hydrogen saturated ASiNR-9 all modes are real, except for some
small imaginary frequencies calculated for the twisting acoustic
mode, TW, near the $\Gamma$ point. This issue was faced earlier
and was attributed to the limitations of the computational
precision.\cite{zno} Thus, the structure is predicted to be
stable. Computational cost of this calculation is very high, so we
were not able to calculate the phonon dispersions for other
ribbons. Nevertheless, all ASiNRs have very similar atomic
configuration, and thus they are also expected to be stable. The
phonon dispersion profile of 2D Ge is similar to that of
Si.\cite{ency} But in Ge structure the acoustic and optic modes
are well separated. Also due to softer bonds the wavenumbers of Ge
structure is halved compared to Si. Thus AGeNRs are also expected
to be stable, whilst exhibiting the mentioned differences.

Phonon densities of states (DOS) of hydrogen saturated 
ASiNR-9 projected to atoms at different locations in the 
nanoribbon are presented in Fig.~\ref{fig:figb}(b). Shown 
is also DOS of the 2D Si honeycomb structure in the same figure.
DOS projected on Si atoms at the center of the nanoribbon 
is very similar to that of the 2D Si. As the width of the 
nanoribbon increases, this similarity is expected
to be enhanced. However, DOS projected on Si atoms at the edges deviate
from that corresponding to 2D Si. Especially, four optical peaks above 
600~$cm^{-1}$ are clearly originating from Si-H bonds at the edges. 
Also modes originating from short Si-Si bonds at the edges cause 
changes in DOS below 600~$cm^{-1}$.

\begin{figure}
\includegraphics[width=8.4cm]{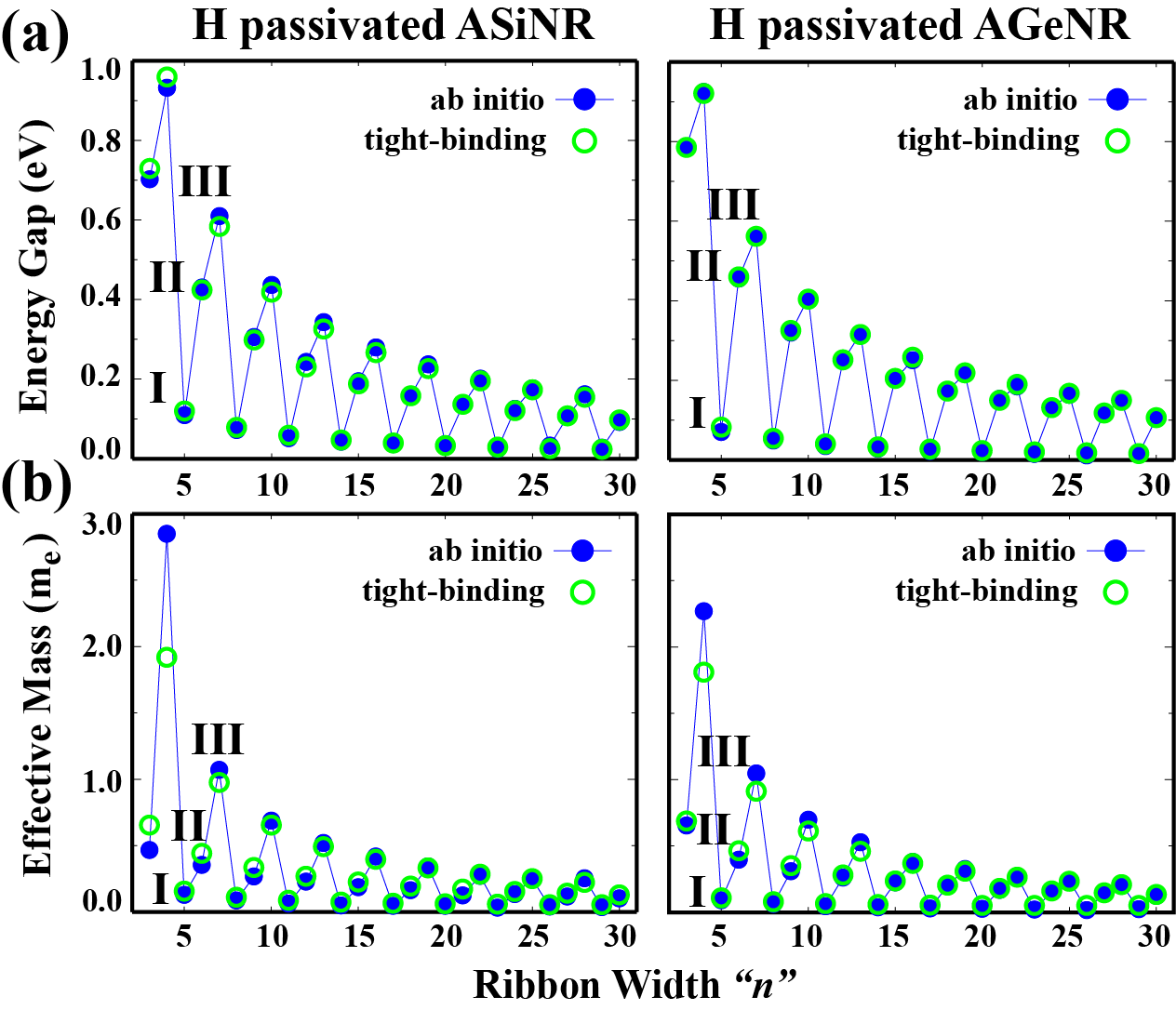}
\caption{Calculated (a) energy gap and (b) effective mass versus
ribbon width, $n$, for hydrogen saturated Si and Ge armchair
nanoribbons. Filled circles indicate the \textit{ab initio}
results while empty circles stand for the results of the tight
binding fitting. The fitting is performed using only the energy
gap data. Parameters found from this fitting was used to generate
the tight binding effective mass data. In each panel three
branches are observed and named in increasing order of band gaps
and effective masses as family I, II, and III.}
\label{fig:figure2}
\end{figure}

\begin{figure*}
\includegraphics[width=15cm]{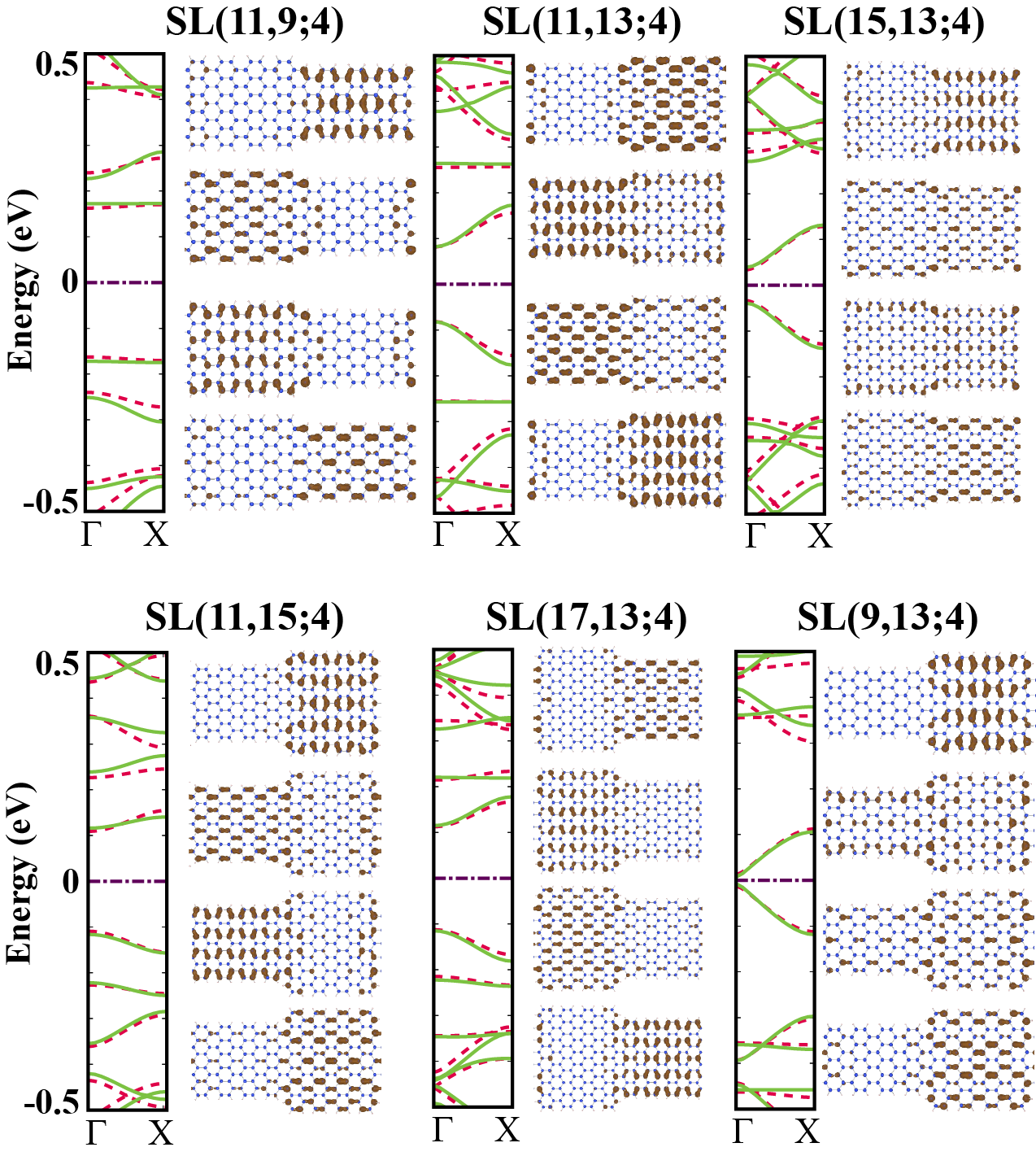}
\caption{Band structure and charge density isosurface profiles for
several superlattice structures. Solid (green) and dashed (red)
lines stand for the \textit{ab initio} and tight binding
calculation results, respectively. Structures are labelled, as defined
in the text, and are given on top of each structure. Top and bottom
panels present the results for superlattice structures with
$\Delta N = 2$ and 4, respectively. Constituent nanoribbons of
superlattices that appear in the same column are members of the
same family. Charge density isosurfaces calculated by \textit{ab
initio} technique around the $\Gamma$ point for two conduction and
two valence band edge states are ordered in the similar manner as
they appear on the energy band structure.} \label{fig:figure3}
\end{figure*}

\section{Electronic Structure}
In this section, we investigate the electronic structure of hydrogen
saturated Si and Ge nanoribbons. Nanoribbons having widths from
$n=3$ to 30 were investigated. Figure~\ref{fig:figure2} presents
the variation of energy gaps and effective masses with the ribbon
width given in terms of $n$. Band gaps are direct and located at
the $\Gamma$. One can see three branches with
decaying profiles originating from the quantum size effect. Here
in ascending order of the band gaps, we would like to name these
branches with widths $n=3k+2$, $3k$, and $3k+1$ as family I, II, and III,
respectively, where $k$ is an integer. This "family behavior" was
also observed in armchair graphene nanoribbons.\cite{gaps} This
trend is explained by foldings of infinite graphene band profile,
which is easily understood by using a tight binding model.
The simplest possible model is to assume that only the first
nearest neighbors interact and have equal hopping parameters with
self-energies set to zero. But this model results in zero band gap
for the members of the family I. Fortunately, this problem is
fixed if a different parameter is used at the edges.\cite{gaps} As
we mentioned in the previous section, the Si-Si (and Ge-Ge) bond
length is apparently smaller at the edges, which implies that edge
bonds are stronger. That is why, defining a different hopping
parameter at the edges reflects the nature of the system better.
Accordingly, we set the nearest neighbor hoping integrals to be 
$t(1+\delta)$ at the edges and $t$ otherwise. 
In our model all self energies were set to zero. 
The tight binding parameters were obtained by 
fitting to the first principles results. Results presented in
Fig.\ref{fig:figure2}~(a) show that the model used here is
successful in reproducing the DFT band gap trends of both Si and
Ge armchair nanoribbons. Since we did not make GW correction to
the band gaps, the nearest neighbor hopping integrals, which are
found to be $t=1.03$~eV for Si and $t=1.05$~eV for Ge, are
expected to have larger values and should be taken as a
qualitative result. However, the relative increase of the hopping
integrals at the edges defined by $\delta$ can be taken as a
quantitative result. We have found $\delta=0.12$ and $\delta=0.08$
for Si and Ge, respectively. Interestingly, the value found for Si
is equal to that of reported value for the armchair graphene
nanoribbons. To sum up, the tight binding parameters determined
for Si and Ge armchair nanoribbons are $t_{Si}=1.03$~eV,
$t_{Ge}=1.05$~eV, $t_{Si,edge}=1.15$~eV and $t_{Ge,edge}=1.13$~eV.

Figure~\ref{fig:figure2} (b) presents the effective masses of the
first conduction band calculated by using the formula
\begin{equation}
m^{*}=\hbar^2 (\frac{\partial^2 E(k)}{\partial k^2})^{-1}
\end{equation}
where $E(k)$ is calculated by using both DFT and tight binding
model mentioned above. Note that, parameters of the tight binding
model are generated by using only the band gap information at the
$\Gamma$ point, but it can reproduce the second order momentum
derivative and thus the effective mass, which is in agreement with
that of \textit{ab initio} calculations. The deviation from this
agreement is seen in the ultimately thin nanoribbons. This is
because, in these nanoribbons the edges affect the rest of the
structure and the tight binding model can not be applied with the
same success. Note that, the effective mass trends are similar to
the band gap trends and show the family behavior. Similar trends
in the effective mass were also observed in the armchair graphene
nanoribbons.\cite{meff}

\begin{figure}
\includegraphics[width=8.4cm]{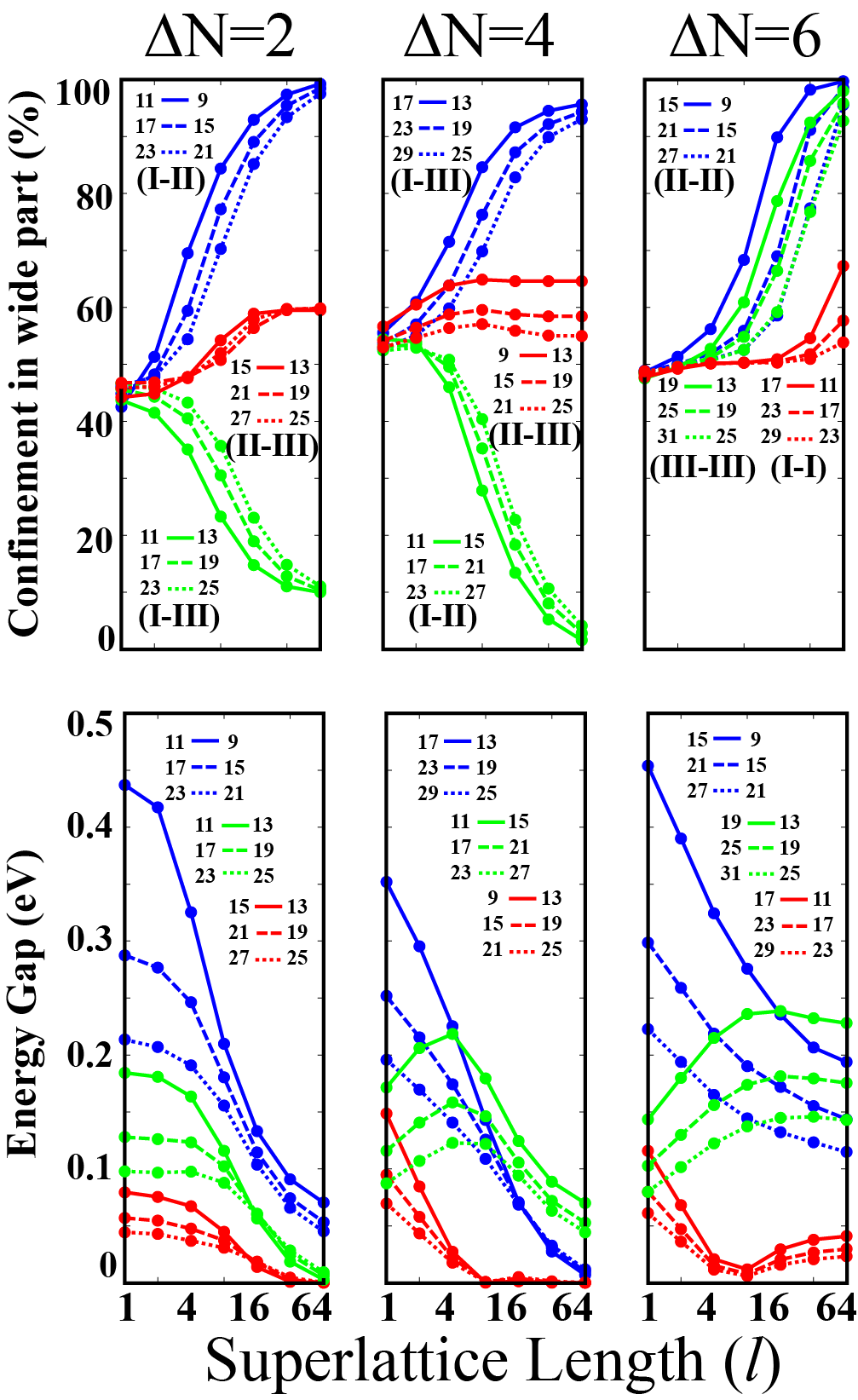}
\caption{Variation of confinements and energy gaps of
superlattices with the lengths of constituent parts. Results are
derived by the tight binding model described in the text. Numbers
labelling the curves correspond to $n_1$ and $n_2$. The lengths,
$l$, of the constituent parts are equal and shown in the
logarithmic scale.} \label{fig:figure4}
\end{figure}

\section{Superlattices}
In this section we investigate the electronic structure of
superlattices formed by periodic junction of ASiNRs which have different
widths. The variation of energy gaps with the ribbon width causes
these structures act as a multiple quantum well and is expected to 
lead to interesting device applications. In the past,
superlattices of 2D GaAs-AlAs or Si-Ge heterostructures have been
extensively studied to realize new generation electronic devices.
Recently, superlattices of graphene have been synthesized
experimentally.\cite{stamp} The transport characteristics and
electron confinement have been revealed.\cite{barbaros_apl} Here
superlattices are formed by the junction of two different
nanoribbons. In this study the lengths of constituent nanoribbons
are taken to be equal. The unequal cases were studied for armchair
nanoribbons of graphene.\cite{hal} We let $l$ unit cell of a
ribbon with the width of $n_1$ to make a perfect junction with $l$
unit cell of a ribbon with the width of $n_2$; once joined these
structures form a supercell having the length of $2l$. Here $n_1$
and $n_2$ are chosen to be odd, so that the ribbons are joined
symmetrically. Then the resulting structure is relaxed and the
lattice constant of the whole structure is determined. This
superlattice structure is labelled by its dimensions as
SL($n_1,n_2;l$). The indices 1 and 2 are arranged in a way that a
ribbon with a width $n_1$ has smaller band gap than a ribbon with
a width $n_2$. The width difference $\Delta N = n_1-n_2$ is
defined for classification purpose.

Figure \ref{fig:figure3} presents the band structure and projected
charge density isosurface plots for sample superlattices having
$\Delta N = 2$ and 4 calculated by DFT. The band profile is nearly
symmetric around the Fermi level and one can easily track the
bonding and antibonding states from the charge density profile. As
seen in the charge density plots, the band edge states of SL($11,9;4$)
and SL($11,13;4$) superlattices are confined in the wide and
narrow part, respectively. This can be explained by taking into 
account the band gaps of constituent ribbons of superlattice
structures. Due to the symmetry between valence and conduction band
edges of the constituent ribbons, the superlattice band line-up is
always normal. As a result, the part having lower gap acts as a
quantum well for electrons and holes. That is why, the electrons
of SL($11,9;4$) and SL($11,13;4$) structures are confined in the
$n=11$ part, which for both superlattices is the part having the
smaller energy gap.

The situation is also similar for the structures having $\Delta N
=~4$. Electrons of SL($17,13;4$) and SL($11,15;4$) are confined,
respectively, in the wide and narrow part, which has the lower
energy gap. Moreover, the superlattices constructed by the
nanoribbons, which are members of the same family, have similar
electronic structure. In Fig.~\ref{fig:figure3} the superlattices
having such common property are shown in the same column. Here one
can see the similarity in the energy band profile of these
structures. Another way to construct such structures is to
increase the width of both constituent parts by $n=6k$. To verify
this, we have calculated the electronic structure of SL($17,15;4$)
and SL($17,19;4$), which yielded in the similar results as that of
SL($11,9;4$) and SL($11,13;4$), except that in former structures
the confinements are less pronounced because the band gap
difference is lower.

We have to mention that, even though the structures SL($9,13;4$)
and SL($15,13;4$) have a considerable band gap difference between
constituent parts, they do not have confinement neither in the
wide nor in the narrow part. This is related to the interface
effects, which will be discussed in a frame of another simple
model.

Calculations for larger structures with \textit{ab initio}
techniques are computationally too expensive. So we have 
used the tight binding model mentioned before. In
Fig.\ref{fig:figure3} one can see that, the tight binding model
can reproduce the band edge profiles over the whole Brillouin
zone. Moreover, the magnitudes of the eigenstates calculated by
tight binding model (not presented here), mimics the projected
charge density profile of a given state. To get a qualitative 
picture of how the large structures behave, the tight
binding model was used.

Figure \ref{fig:figure4} presents the energy gap and the band edge
confinement strength trends generated by the tight binding model
for the superlattices having $n = 9-31$, $\Delta N = 2-6$ and 
$l = 1-64$. The magnitude squares, $|\Psi|^2$, of conduction and valence
band edge eigenfunctions generated by the tight binding model are
equal. The confinement percentage in the narrow (wide) part is
defined as the sum of the magnitude squares of band edge
eigenfunctions in the narrow (wide) part multiplied by 100 \%,
noting that the overall sum of the magnitude squares is normalized
to 1. In Fig.~\ref{fig:figure4} we see that, in agreement with the
\textit{ab initio} results presented in Fig.~\ref{fig:figure3}, the
band edge states are confined in the narrow part for SL($11,15;l$)
and SL($11,13;l$) structures and in the wide part for SL($11,9;l$)
and SL($17,13;l$) structures. These confinements are enhanced as
the lengths $l$ of the segments increase, because the strength of
the well and the barrier increase. The trends shown in
Fig.~\ref{fig:figure4} also confirm the statement that the
superlattices composed of the nanoribbons from the same family
behave similarly. For comparison, in each panel of the
Fig.~\ref{fig:figure4} structures having such similarity are shown
in the same style. Also the corresponding family names are
presented under the figure labels. The similarity can be seen in
both confinement and the energy gap trends. For structures having
$\Delta N = 6$, the band edge states are always confined at the
wide part. This is because, the band gap for structures having
$\Delta N = 6$ is lower in the wide part, so the wide part acts as
a quantum well.

In Fig.~\ref{fig:figure4}, one can see that there is no strong
confinement in the wide or narrow part of SL($9,13;l$) and
SL($15,13;l$) structures. This is also seen in the \textit{ab
initio} results presented in Fig.~\ref{fig:figure3}. Moreover, the
case is also true for the superlattices composed of the
nanoribbons from the same family. Nevertheless, this can not be
explained by the band gap difference of the constituent
nanoribbons, because there are superlattices composed of the
nanoribbons having similar band gap difference but show strong
confinement patterns. Actually, plotting the linear charge density
of former structures along the periodic direction results in
decaying profiles in both narrow and wide parts. Thus for these
structures, the band edge states are localized at the interfaces.
This means that, the interface acts as a quantum well in these
structures.

\begin{figure}
\includegraphics[width=8.4cm]{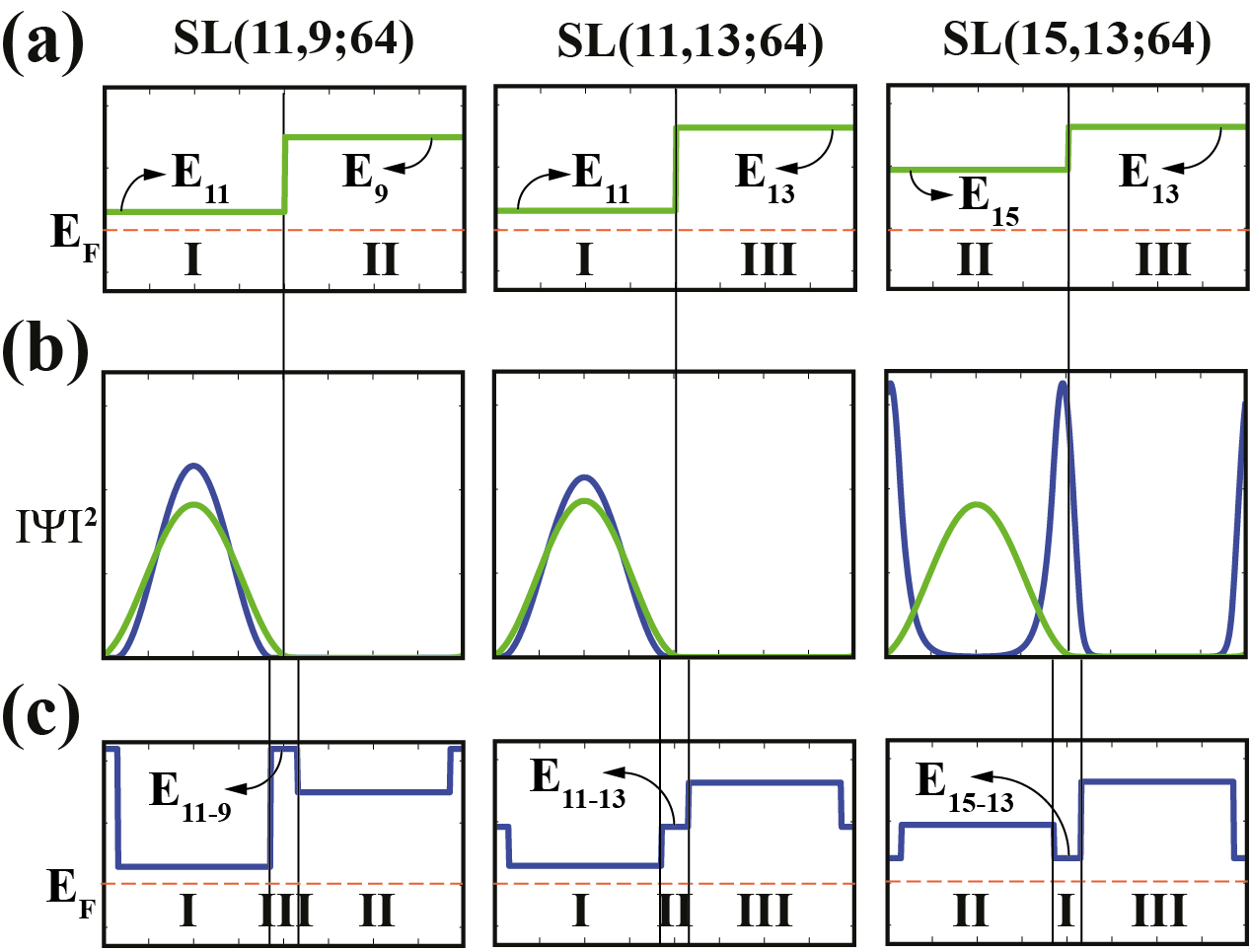}
\caption{Model representation of three different superlattice
structures. (a) Conduction band edge profile when interface
effects are not included. Here the height of the potential in each
side is determined by E$_n$, which is the difference between the
conduction band edge minimum and Fermi level of a nanoribbon with
a width $n$. Fermi levels of each structure are shown by the
dashed lines. The family name corresponding to each side is
written under these lines. (b) Magnitude squares ($|\Psi|^{2}$) of
solutions. Dark (blue) and light (green) lines represent solutions
for the cases where interface effects are included and are not
included, respectively. The thin line joining the panels (a) and
(b) corresponds to the geometric interface of the superlattices.
(c) Conduction band edge profile when interface effects are
included. Here the height of the potential at the interface is
determined by E$_{n_1-n_2}$, which is the difference between the
conduction band edge minimum and Fermi level of SL($n_1,n_2;1$)
structure. The interfaces act as if they are composed of the
families I, II, or III as written in that region.}
\label{fig:figure5}
\end{figure}

To consolidate the effects of the interface, we have to make an
estimation about its effective energy gap and mass. For this
purpose, we have chosen the energy gap and the effective mass of
SL($n_1,n_2;1$) structure to represent the effect of the interface
between the nanoribbons with widths $n_1$ and $n_2$. This
information was used in construction of a very simple 1D 
model, which qualitatively explains the confinement
trends mentioned so far. Here the band edge state of a
superlattice structure SL($n_1,n_2;l$) is modelled as a single
electronic state which is under the influence of a periodic
potential. The height of this potential is taken to be equal to
the difference between the conduction band minimum and Fermi level
of each region. The resulting profiles are shown for three
different superlattice structures in Fig.~\ref{fig:figure5} (a).
Here the interface effects are not included and the length
parameter of each part is set to 64 times the length of a
unit cell. Also the mass of electrons in each segment is modelled
by the effective mass of a nanoribbon with a width $n$ of that
segment. The magnitude square of the numerical solution of this
system is presented with a light (green) line in the
Fig.~\ref{fig:figure5} (b). To include the interface effects, the
potential height of an arbitrary small region at the interface is
changed to the difference between conduction band minimum and
Fermi level of SL($n_1,n_2;1$) structure. The dark (blue) lines in
the Fig.~\ref{fig:figure5} (c) represent the new effective
potential profiles. Solutions of these systems are given by dark
(blue) lines in the Fig.~\ref{fig:figure5} (b).

Examining the changes induced by inclusion of the interface
effects given in Fig.~\ref{fig:figure5} (b) one can deduce that in
SL($11,9;64$) and SL($11,13;64$) structures the interface having
large effective mass acts as a barrier which slightly enhances the
confinement strength. In parallel to this, one can see in
Fig.~\ref{fig:figure3} that the charge densities are more
localized in SL($11,9;4$) structure than in SL($11,13;4$)
structure. Also the enhanced confinement of SL($11,9;l$) structure
is seen in trends presented in Fig.~\ref{fig:figure4}. In
SL($15,13;64$) structure, however, the effect of the interface is
dramatic. Since the effective energy gap of the interface region
is smaller than that of the narrow and wide parts, it acts as a
quantum well. As a result, we have a charge density confined in
the interface region and decaying in the narrow and wide parts.
The decay is sharper in the region where the effective potential
is deeper and the mass is higher. This profile is similar to the
one found by tight binding model.

Trends found in this section can be summarized by very simple
arguments; (i) if the superlattices are formed by two different
families then the interface acts as if it is formed by a family
different from both (see Fig.~\ref{fig:figure5} (c)); (ii) in this
case each part acts differently and electrons are confined in the
part acting as the family I, which has the lowest band gap; (iii)
if both sides are composed of the same family, then the electrons
are confined at the wider part; (iv) confinements are enhanced
by increasing the lengths of the constituent parts.

\section{Conclusion}

We have investigated atomic structure, stability and electronic
properties of armchair silicon and germanium nanoribbons by performing first
principles calculations. The edges of bare armchair nanoribbons 
are reconstructed. The reconstruction is, however, removed by the 
hydrogen passivation of the edge atoms. It was shown that, these nanoribbons
exhibit, so called, family behavior which was explored also in
graphene nanoribbons. A simple tight binding model proposed for
graphene nanoribbons was shown to work very well for silicon and
germanium nanoribbons. Superlattices formed by periodic modulation
of silicon nanoribbon widths were also investigated. Modulation of
widths in the real space attributes these structures multiple
quantum well properties. Specific electronic states are confined
in these superlattice structures. Confinements increase with
increasing the lengths of constituent parts of the superlattice.
In general, the band edge states are confined in the part of
superlattice which have the smallest band gap by itself. This part
can be narrow, wide or the interface part of the superlattice.
Superlattices composed of the nanoribbons which are members of the
same family have similar electronic structure. Confinement
patterns can be explained by inclusion of the interface effects in
a frame of a simple 1D effective potential model. It is demonstrated 
that Si and Ge nanoribbons and superlattices constructed therefrom 
display features, which may be exploited in future nanodevices.

\end{document}